\documentclass[useAMS,usenatbib]{mn2e}
\usepackage{times,graphicx,aas_macros}
\usepackage{float}
\usepackage{ifthen}
\usepackage{times}
\usepackage{natbib}
\usepackage{rotating}
\usepackage{upgreek}
\DeclareGraphicsExtensions{.pdf,.png,.jpg}

\begin{document}

\title[S2CLS: 450$\mu$m number counts and the CIB]{The SCUBA--2 Cosmology
Legacy Survey: blank-field number counts of 450$\upmu$m-selected galaxies
and their contribution to the cosmic infrared background}

\author[J.\ E.\ Geach et al.]{\parbox[h]{\textwidth}{J.\ E.\ Geach$^1$\thanks{E-mail: jimgeach@physics.mcgill.ca},
E.\ L.\ Chapin$^{2,3,4}$, K.\ E.\ K.\ Coppin$^1$, J.\ S.\ Dunlop$^5$,
M.\ Halpern$^2$, Ian\ Smail$^6$, P.\ van der Werf$^7$, S.\ Serjeant$^8$,
D.\ Farrah$^9$, I.\ Roseboom$^5$, T.\ Targett$^5$, V.\ Arumugam$^5$, V.\
Asboth$^3$, A.\ Blain$^{10}$, A.\ Chrysostomou$^{3,11}$,  C.\ Clarke$^{12}$, R.\ J.\
Ivison$^{5,13}$, S.\ L.\ Jones$^{10}$,\ A.\ Karim$^6$, T.\ Mackenzie$^2$, R.\ Meijerink$^{7,14}$, M.\ J.\
Micha{\l}owski$^5$, D.\ Scott$^2$, J.\ Simpson$^6$, A.\ M.\ Swinbank$^6$, D.\ Alexander$^6$, O.\
Almaini$^{15}$, I.\ Aretxaga$^{16}$, P.\ Best$^5$, S.\ Chapman$^{17}$,
D.\ L.\ Clements$^{18}$, C.\ Conselice$^{15}$, A.\ L.\ R.\ Danielson$^6$, S.\ Eales$^{19}$,
A.\ C.\ Edge$^6$, A.\ Gibb$^3$, D.\ Hughes$^{16}$, T.\ Jenness$^{3}$,
 K.\ K.\ Knudsen$^{20}$, C.\ Lacey$^6$, G.\
Marsden$^2$, R.\ McMahon$^{21}$, S.\ Oliver$^{12}$, M.\ J.\ Page$^{22}$, J.\ A.\ Peacock$^5$,
D.\ Rigopoulou$^{23,24}$, E.\ I.\ Robson$^{13}$, M.\ Spaans$^{14}$, J.\
Stevens$^{11}$,  T.\ M.\ A.\ Webb$^1$, C.\ Willott$^{25}$,
C.\ D.\ Wilson$^{26}$, M.\ Zemcov$^{27}$}
\vspace*{12pt}\\ \noindent $^{1}$Department of Physics, Ernest Rutherford
Building, 3600 rue University, McGill University, Montr\'eal, QC, H3A 2T8,
Canada\\
\noindent $^{2}$ Department of Physics \& Astronomy, University of British Columbia, 6224 Agricultural Road, Vancouver, BC, V6T 1Z1, Canada\\
\noindent $^{3}$ Joint Astronomy Centre 660 N. A'ohoku Place University Park
Hilo, Hawaii 96720, USA\\
\noindent $^{4}$ XMM SOC, ESAC, Apartado 78, 28691 Villanueva de la Canada, Madrid, Spain\\
\noindent $^{5}$ Institute for Astronomy, University of Edinburgh, Royal Observatory, Blackford Hill, Edinburgh EH9 3HJ\\
\noindent $^{6}$ Institute for Computational Cosmology, Department of Physics, Durham University, South Road, Durham, DH1 3LE\\
\noindent $^{7}$ Leiden Observatory, Leiden University, P.O. box 9513, 2300 RA Leiden, The Netherlands\\
\noindent $^{8}$ Robert Hooke Building, Department of Physical Sciences, The Open University, Milton Keynes, MK7 6AA\\
\noindent $^{9}$ Virginia Polytechnic Institute \& State University 
Department of Physics, MC 0435, 910 Drillfield Drive, Blacksburg, VA 24061, USA\\
\noindent $^{10}$ Department of Physics \& Astronomy, University of Leicester,
University Road, Leicester, LE1 7RH\\
\noindent $^{11}$ Centre for Astrophysics Research, Science \& Technology
Research Institute, University of Hertfordshire, Hatfield, AL10 9AB\\
\noindent $^{12}$ Astronomy Centre, Department of Physics and Astronomy,
University of Sussex, Brighton BN1 9QH\\
\noindent $^{13}$ UK Astronomy Technology Centre, Royal Observatory, Blackford
Hill, Edinburgh EH9 3HJ\\
\noindent $^{14}$ Kapteyn Institute, University of Groningen, P.O. Box 800,
9700 AV Groningen, The Netherlands\\
\noindent $^{15}$ School of Physics and Astronomy, University of Nottingham,
University Park, Nottingham, NG9 2RD\\
\noindent $^{16}$ Instituto Nacional de Astrof\'isica \'Optica y
Electr\'onica, Calle Luis Enrique Erro No. 1, Sta. Ma. Tonantzintla, Puebla,
M\'exico\\
\noindent $^{17}$ Department of Physics and Atmospheric Science, Dalhousie
University Halifax, NS, B3H 3J5, Canada\\
\noindent $^{18}$ Astrophysics Group, Imperial College London, Blackett
Laboratory, Prince Consort Road, London, SW7 2AZ\\
\noindent $^{19}$ Cardiff School of Physics and Astronomy, Cardiff University,
Queens Buildings, The Parade, Cardiff, CF24 3AA\\
\noindent $^{20}$ Department of Earth and Space Science, Chalmers University
of Technology, Onsala Space Observatory, SE-43992 Onsala, Sweden\\
\noindent $^{21}$ Institute of Astronomy, University of Cambridge, Madingley Road, Cambridge, CB3 OHA \\
\noindent $^{22}$ Mullard Space Science Laboratory, University College London, Holmbury St Mary Dorking, Surrey RH5 6NT\\
\noindent $^{23}$ Department of Physics, University of Oxford, Keble Road, Oxford, OX1 3RH\\
 \noindent $^{24}$ Space Science \& Technology Department, Rutherford Appleton Laboratory, Chilton, Didcot, Oxfordshire, OX11 0QX\\
\noindent $^{25}$ Canadian Astronomy Data Centre, National Research Council
Canada, 5071 West Saanich Road, Victoria, BC, V9E 2E7, Canada\\
\noindent $^{26}$ Department of Physics and Astronomy, McMaster University
Hamilton, ON, L8S 4M1, Canada\\
\noindent $^{27}$ Astronomy Department, California Institute of Technology, MC
367-17 1200 East California Blvd., Pasadena, CA 91125, USA 
}

\maketitle 

\label{firstpage}

\clearpage

\begin{abstract}The first deep blank-field 450$\mu$m map
($1\sigma\approx1.3$\,mJy) from the SCUBA--2 Cosmology Legacy Survey (S2CLS),
conducted with the James Clerk Maxwell Telescope (JCMT) is presented. Our map
covers 140\,arcmin$^2$ of the Cosmological Evolution Survey (COSMOS) field, in
the footprint of the {\it Hubble Space Telescope} ({\it HST}) Cosmic Assembly
Near-Infrared Deep Extragalactic Legacy Survey (CANDELS). Using 60
submillimetre galaxies (SMGs) detected at $\geq$3.75$\sigma$, we evaluate the
number counts of 450$\mu$m-selected galaxies with flux densities
$S_{450}>5$\,mJy. The 8$''$ JCMT beam and high sensitivity of SCUBA--2 now
make it possible to directly resolve a larger fraction of the cosmic infrared
background (CIB, peaking at $\lambda\sim200$$\mu$m) into the individual
galaxies responsible for its emission than has previously been possible at
this wavelength. At $S_{450}>5$\,mJy we resolve
$(7.4\pm0.7)\times10^{-2}$\,MJy\,sr$^{-1}$ of the CIB at 450$\mu$m (equivalent
to $16\pm7$\% of the absolute brightness measured by the {\it Cosmic
Background Explorer} at this wavelength) into point sources. A further
$\sim$40\% of the CIB can be recovered through a statistical stack of 24$\mu$m
emitters in this field, indicating that the majority ($\approx$60\%) of the
CIB at 450$\mu$m is emitted by galaxies with $S_{450}>2$\,mJy. The average
redshift of 450$\mu$m emitters identified with an optical/near-infrared
counterpart is estimated to be $\langle z \rangle=1.3$, implying that the
galaxies in the sample are in the ultraluminous class ($L_{\rm IR}\approx
1.1\times10^{12}{\rm L_\odot}$). If the galaxies contributing to the
statistical stack lie at similar redshifts, then the majority of the CIB at
450$\mu$m is emitted by galaxies in the LIRG class with $L_{\rm
IR}>3.6\times10^{11}{\rm L_\odot}$.\end{abstract}


\section{Introduction}

\begin{figure*}\centerline{
\includegraphics[width=0.5\textwidth]{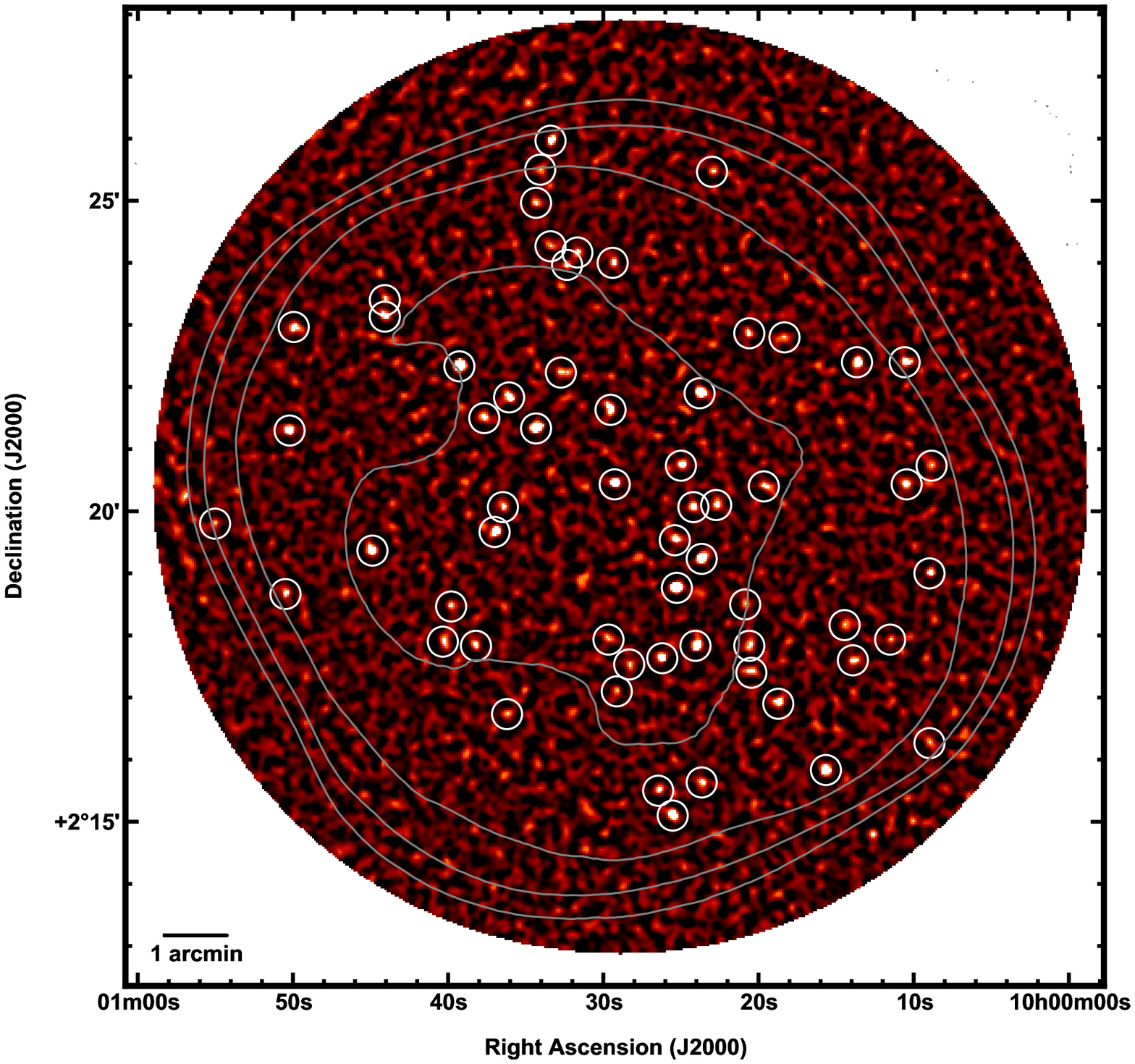}\includegraphics[width=0.5\textwidth]{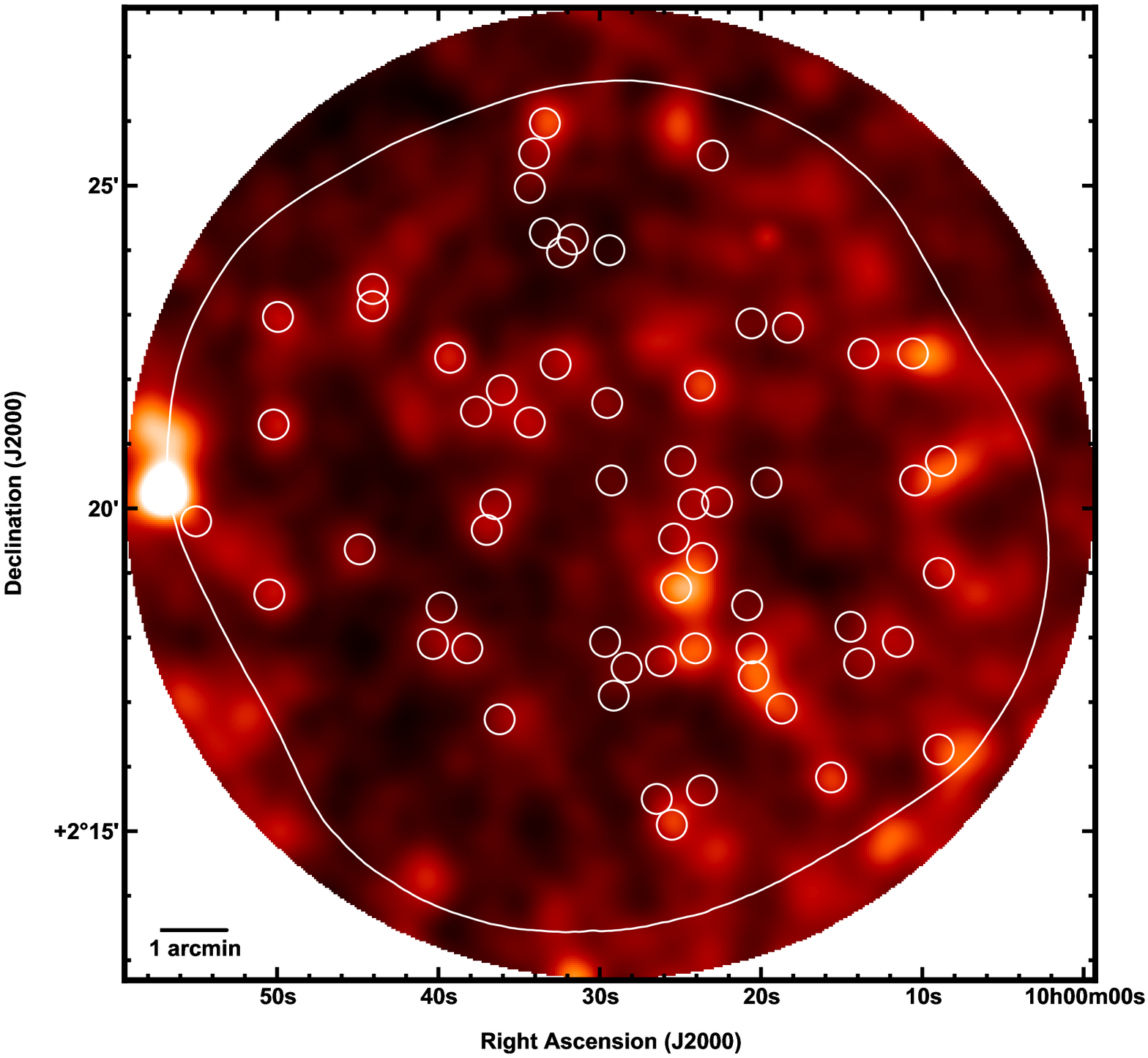}}
\caption{(left) SCUBA--2 450$\mu$m signal-to-noise ratio map of the
COSMOS/CANDELS field. The map has been scaled to emphasize the visibility of
60 sources detected at $>$3.75$\sigma$ significance (circled). The grey
contours show the variation in the noise level, and are at $\sigma_{\rm
450}=2,3,4,5$\,mJy\,beam$^{-1}$ (the solid angle bounded by the
$\sigma_{450}=5$\,mJy\,beam$^{-1}$ contour is $\Omega\approx140$\,arcmin$^2$).
(right) {\it Herschel}-SPIRE 500$\mu$m image of the same region, from the
HerMES survey (Oliver et al.\ 2012). This map has been slightly smoothed with
a Gaussian kernel to improve presentation. We show the limiting
$\sigma_{450}=5$\,mJy\,beam$^{-1}$ contour used for 450$\mu$m detection in the
SCUBA--2 map and the position of the same galaxies in the left panel. This
illustrates the ability of SCUBA--2 to push below the {\it Herschel} confusion
limit at similar wavelengths, resolving confused emission into individual
galaxies. } \end{figure*}

Fifteen years have passed since the first `submillimetre galaxies' (SMGs) were
discovered (Smail, Ivison \& Blain\ 1997; Barger et al.\ 1998; Hughes et al.\
1998), a high-redshift population ($z\sim2-3$, Chapman et al.\ 2005; Aretxaga
et al.\ 2007; Wardlow et al.\ 2011) with ultraluminous ($10^{12}{\rm
L_\odot}$) levels of bolometric emission, the bulk of which is emitted in the
far-infrared (FIR) and redshifted to submillimetre wavelengths at $z>1$. The
power of submillimetre surveys for exploring the formation phase of massive
galaxies was recognised before their discovery (e.g.\ Blain \& Longair\ 1993;
Dunlop et al.\,1994), and since their discovery, their importance as a
cosmologically significant population has been established by many studies
(e.g.\ Smail et al.\ 2002; Dunlop et al.\ 2004; Ivison et al.\ 2000, 2005,
2010; Coppin et al.\ 2008; Micha{\l}owski et al.\ 2010; Hainline et al.\ 2011;
Hickox et al.\ 2012; and see Dunlop et al.\ 2011 for a review). As such, SMGs
provide challenging tests for models of galaxy formation, both in detailed
`zoomed' simulations as well as in cosmological theatres (Baugh et al.\ 2005;
Dav\'e et al.\ 2010). However, our view of the SMG population remains
incomplete.

In ground-based work, the majority of SMGs have -- so-far -- mainly been
selected in the 850$\mu$m or 1\,mm atmospheric windows (e.g.\ Coppin et al.\
2006; Weiss et al.\ 2009; Austermann et al.\ 2010; Scott et al.\ 2010), but
this is far removed from the peak of the Cosmic infrared background (CIB),
which is at $\lambda\sim200$$\mu$m (Fixsen et al.\ 1998). The next available
window closer to the CIB peak is at 450$\mu$m, but the transmission of this
window is just at best 50\% of the 850$\mu$m window, making 450$\mu$m SMG
surveys challenging from ground-based sites. Submillimetre surveys working
closer to the CIB peak are essential if we are to identify the galaxies
responsible for its emission; the $S_{450}/S_{850}$ colours of sources
identified in the very deepest (lensing assisted) submillimetre surveys (Blain
et al.\ 1999; Knudsen et al.\ 2008) suggests that these sources contribute
less than half of the CIB at 450$\mu$m (and therefore even less at the actual
peak).

The Balloon-borne Large Aperture submillimetre Telescope (BLAST, Pascale et
al.\ 2008) made progress by conducting a low resolution submillimetre survey
from the stratosphere at 250, 350 and 500$\mu$m (Pascale et al.\ 2008; Devlin
et al.\ 2009; Glenn et 2010). This work was taken forward by the {\it Herschel
Space Observatory}, which carries an instrument that images in the same
wavelength ranges as BLAST (the Spectral and Photometric Imaging Receiver;
SPIRE), and has mapped hundreds of square degrees of the sky at 250--500$\mu$m
in a combination of panoramic and deep cosmological surveys (Eales et al.\
2010, Oliver et al.\ 2010, 2012). However, the low resolution and high
confusion limits of {\it Herschel} ({\sc fwhm}$\sim$0.5$'$ at 500$\mu$m,
$\sigma_{\rm con}\approx7$\,mJy\,beam$^{-1}$, Nguyen et al.\ 2010) limit the
fraction of the CIB that can be directly resolved, with 15\% resolved into
individual galaxies at 250$\mu$m and 6\% at 500$\mu$m (Clements et al.\ 2010;
Glenn et al.\ 2010; B\'ethermin et al.\ 2012a). Thus, there remains work to be
done in identifying the galaxies that emit the CIB, and thus finally complete
the census of dust-obscured activity in the Universe and its role in galaxy
evolution.

Advances in submillimetre imaging technology are just now allowing us to take
up the search once more, taking advantage of higher resolution possible with
large terrestrial telescopes, and improved sensitivity and mapping capability
in submillimetre detector arrays. The SCUBA--2 camera is the state-of-the-art
in submillimetre wide-field instrumentation (Holland et al.\ 2006). The
camera, now mounted on the 15\,m James Clerk Maxwell Telescope (JCMT),
consists of 5000 pixels in both 450$\mu$m and 850$\mu$m detector arrays with
an $8'\times8'$ field-of-view (16$\times$ that of its predecessor, SCUBA). The
increase in pixel number is the reward of developments in submillimetre
detector technology; SCUBA--2 utilizes superconducting transition edge sensors
(TES) to detect submillimetre photons, with multiplexed superconducting
quantum interference device (SQUID) amplifiers handling read-out, analogous to
an optical CCD. SCUBA--2 offers the capability to efficiently map large
(degree-scale) areas, and has the sensitivity to simultaneously achieve deep
(confusion limited) maps at both 450$\mu$m and 850$\mu$m. At 450$\mu$m, the
resolution attainable with the JCMT is a factor $\sim$5$\times$ finer than the
500$\mu$m resolution of {\it Herschel}, and the confusion limit is
$\sim$7$\times$ fainter.

Here we present results from early science observations of one of the seven
components of the JCMT Legacy
Survey\footnote{http://www.jach.hawaii.edu/JCMT/surveys}: the SCUBA--2
Cosmology Legacy Survey
(S2CLS)\footnote{http://www.jach.hawaii.edu/JCMT/surveys/Cosmology.html}. The
goal of the S2CLS is to fully exploit SCUBA--2's mapping capabilities for the
purpose of exploring the high redshift Universe. The S2CLS will cover several
well-studied extragalactic `legacy' fields, including the United Kingdom
Infrared Deep Sky Survey Ultra Deep Survey field (UDS), the Cosmological
Evolution field (COSMOS), the Extended Groth Strip, and the Great
Observatories Origins Deep Survey (North) fields. We present the deepest
blank-field map at 450$\mu$m yet produced (in the COSMOS field), and measure
the flux distribution and abundance of the extragalactic sources revealed
within it. In \S2 we describe the observations and data reduction technique,
in \S3 we calculate the 450$\mu$m number counts and evaluate the contribution
to the CIB at 450$\mu$m. We briefly discuss and summarize our findings in \S4
\& \S5.

\section{Observations and data reduction}

Observations were conducted in Band 1 weather conditions ($\tau_{\rm
225\,GHz}<0.05$) over 22 nights between 23$^{\rm rd}$ January and 20$^{\rm
th}$ May 2012 totalling 50 hours of on-sky integration. The mapping centre of
the SCUBA--2 COSMOS/CANDELS field is $\alpha=$\,10$^{\rm h}$\ 10$^{\rm m}$\
29.8$^{\rm s}$, $\delta=$\,02$^{\circ}$\ 15${'}$\ 01.6${''}$, chosen to be in
the footprint of the {\it Hubble Space Telescope} CANDELS (Grogin et al.\
2011; Koekemoer et al.\ 2011)\footnote{Cosmic Assembly Near-infrared Deep Extragalactic Legacy Survey,
{ http://candels.ucolick.org/}}. A standard 3\,arcmin diameter `daisy' mapping
pattern was used, which keeps the pointing centre on one of the four SCUBA--2
sub-arrays at all times during exposure.

\begin{figure}
\centerline{\includegraphics[width=0.38\textwidth,angle=-90]{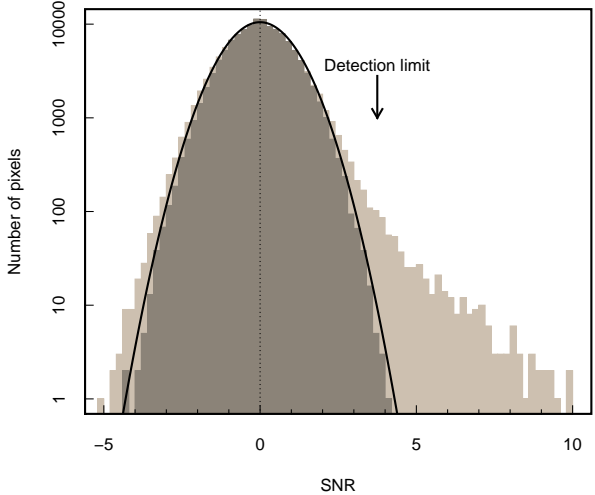}}
\caption{Histogram of values in the SCUBA--2 450$\mu$m signal-to-noise ratio
map (Fig.\,\ 1), indicating the characteristic positive tail due to the
presence of real astronomical sources. The solid line is a Gaussian centred at
zero with a width of $\sigma=1$, and the darker shaded histogram shows the
histogram of pixel values in a map constructed by inverting a random 50\% of
the input scans; we use this for simulations of completeness, described in
\S2.3. Our detection limit is chosen to be $\sigma=3.75$, which yields a
reasonably complete and reliable catalogue (see \S2.3). Note that the `real'
map noise distribution is slightly wider than expected for pure Gaussian
noise; this is due to slight `ringing' around bright sources after convolution
with the beam.} \end{figure}

\subsection{Map making}

Individual 30\,min scans are reduced using the dynamic iterative map-maker of
the {\sc smurf} package (Jenness et al.\ 2011; Chapin et al.\ 2012 in prep).
Raw data are first flat-fielded using ramps bracketing every science
observation, scaling the data to units of pW. The signal recorded by each
bolometer is then assumed to be a linear combination of: (a) a common mode
signal dominated by atmospheric water and ambient thermal emission; (b) the
astronomical signal (attenuated by atmospheric extinction); and finally (c) a
noise term, taken to be the combination of any additional signal not accounted
for by (a) and (b). The dynamic iterative map maker attempts to solve for
these model components, refining the model until convergence is met, an
acceptable tolerance has been reached, or a fixed number of iterations has
been exhausted (in this case, 20). This culminates in time-streams for each
bolometer that should contain only the astronomical signal, corrected for
extinction, plus noise. The signal from each bolometer's time stream is then
re-gridded onto a map, according to the scan pattern, with the contribution to
a given pixel weighted according to its time-domain variance (which is also
used to estimate the $\chi^2$ tolerance in the fit derived by the map maker).

The sky opacity at JCMT has been obtained by fitting extinction models to
hundreds of standard calibrators observed since the commissioning of SCUBA--2
(Dempsey et al.\ 2012). The optical depth in the 450$\mu$m band was found to
scale with the Caltech Submillimetre Observatory (CSO) 225\,GHz optical depth
as: $\tau_{450}=26.0(\tau_{225}-0.0196)$. Note that this scaling is slightly
different from the original SCUBA relations (see Archibald et al.\ 2002;
Dempsey et al.\ 2012).

Filtering of the time-series is performed in the frequency domain, with
band-pass filters equivalent to angular scales of $2''<\theta<120''$ (i.e.\
frequencies of $f=v/\theta$, where $v$ is the scan speed). The reduction also
includes the usual filtering steps of spike removal ($>$10$\sigma$ deviations
in a moving boxcar) and DC step corrections. Throughout the iterative map
making process, bad bolometers (those significantly deviating from the model)
are flagged and do not contribute to the final map. Maps from independent
scans are co-added in an optimal stack using the variance of the data
contributing to each pixel to weight spatially aligned pixels. Finally, since
we are interested in (generally faint) extragalactic point sources, we apply a
beam matched filter to improve point source detectability, resulting in a map
that is convolved with an estimate of the 450$\mu$m beam. The average exposure
time over the nominal 3 arcminute daisy mapping region (in practice there is
usable data beyond this) is approximately $10$\,ksec per $2''\times2''$ pixel.

\subsection{Flux calibration}

The flux calibration of SCUBA--2 data has been examined by analysing all flux
calibration observations since Summer 2011 until the date of observation. The
derived beam-matched flux conversion factor (FCF) has been found to be
reasonably stable over this period, and the average FCFs agree (within error)
to those derived from the subset of standard calibrators observed on the
nights of the observations presented here. Therefore we have adopted the
canonical calibration of ${\rm
FCF}_{450}=540\pm65$\,Jy\,beam$^{-1}$\,pW$^{-1}$ here. A correction of
$\sim$10\% is included in order to compensate for flux lost due to filtering
in the blank-field map. This is estimated by inserting a bright Gaussian point
source into the time stream of each observation to measure the response of the
model source to filtering.

\subsection{Maps and source detection}

We present the 450$\mu$m signal-to-noise ratio map of the COSMOS/CANDELS field
in Fig.\,\,1. For comparison, we also show a {\it Herschel} SPIRE 500$\mu$m
map of the same region to illustrate the gain in resolution that JCMT/SCUBA--2
offers at similar wavelengths\footnote{The {\it Herschel} map was made from
the Level 2.5 processed data products downloaded from the public {\it
Herschel} Science Archive. The data were co-added with sky coverage used as an
estimator for image noise level, and re-binned into the SCUBA--2 image
reference frame, using nearest-neighbour sampling. The 1$\sigma$ noise level
of this SPIRE map (including confusion) is 6.2\,mJy\,beam$^{-1}$}. The
450$\mu$m map has a radially varying sensitivity, which is nearly uniform over
the central 3$'$ (the nominal mapping area) and smoothly increases in the
radial direction as the effective exposure time decreases for pixels at the
edge of the scan pattern, which have fewer bolometers contributing to the
accumulated exposure. The total area of the map considered for source
extraction is 140\,arcmin$^2$, where the rms noise is below
5\,mJy\,beam$^{-1}$. A histogram of pixel values in the
$\sigma_{450}\leq5$\,mJy\,beam$^{-1}$ region is shown in Fig.\,\ 2.

To identify extragalactic point sources, we search for pixels in the (beam
convolved) signal-to-noise ratio map with values $>$$\Sigma_{\rm thresh}$. If
a peak is found, we record the peak-pixel sky co-ordinate, flux density and
noise, mask-out a circular region equivalent to $\simeq$1.5$\times$ the size
of the 8$''$ beam at 450$\mu$m, reduce $\Sigma_{\rm thresh}$ by a small amount
and then repeat the search. The floor value, below which we no longer trust
the reality of `detections' is chosen to be the signal-to-noise level at which
the contamination rate due to false detections (expected from pure Gaussian
noise) exceeds 5\%, corresponding to a significance of $\sigma$$\approx$3.75. We detect 60 discrete point sources in this way, and these are identified in Fig.\,\ 1. Note that the map is far from confused, with an average source density equivalent to $6\times10^{-3}$\,beam$^{-1}$. We project that the confusion limit is at $\sim$1\,mJy\,beam$^{-1}$

Completeness is estimated by injecting a noise model with artificial point
sources. To create maps with no astronomical sources but approximately the
same noise properties of the real map, we generate jackknife realisations of
the map where, in each fake map, a random half of individual scans have their
signal inverted before co-addition (e.g.\ Weiss et al.\ 2009). Fig.\,2 shows
the equivalent histogram of signal-to-noise ratio values in the jackknife map,
which demonstrates the clean removal of astronomical sources, and the
similarity with pure Gaussian noise. The recovery rate of sources as a
function of input flux and local noise gives the completeness function: $10^5$
fake sources in batches of 10 are inserted into the jackknife map, where each
source selected from a uniform flux distribution $1 < (S_{450}/{\rm mJy}) <
40$. The 2-dimensional completeness function is shown in Fig.\,\ 3.

In addition to the completeness correction, this technique allows us to
estimate the noise-dependent flux boosting that occurs for sources with true
fluxes close to the noise limit of the map, and so we can construct an
equivalent `surface' in the noise--(measured) flux plane that can be use to
de-boost the fluxes measured for point sources in the real map (Table\ 1). The
typical de-boosting correction is $\mathcal{B}<$10\%. Finally, the source detection
algorithm is applied to each of the jackknife maps with no fake sources
injected in order to evaluate the false positive rate, which we find to be
5\%, in agreement with the false detection rate expected for a map of this
size assuming fluctuations from pure Gaussian noise.

A test for any bias in the recovery and correction of the source counts was
performed in the following way. We populated the jackknife maps with a model
source count model (B\'ethermin et al.\ 2012b) down to a flux limit of
$S_{450}=0.01$\,mJy. Sources were then extracted in exactly the same manner as
the real data and completeness and flux boosting corrected as described above
and then compared to the input distribution. This procedure was repeated 100
times and the average recovered source counts compared to the input model. The
recovered differential and cumulative number counts were found to be
consistent with the input number count realisations, indicating that our
source detection and completeness corrections are not significantly biased.

\section{Analysis}

\subsection{Number counts of 450${\bf \mu}$m emitters}

In Table\ 1 and Fig.\,\ 4 we present the number counts at 450$\mu$m, corrected
for flux boosting and incompleteness. The differential counts are
well-described by a Schechter function:

\begin{equation}
	\frac{dN}{dS} = \left(\frac{N'}{S'}\right) \left(\frac{S}{S'}\right)^{1-\alpha}\exp\left(\frac{N'}{S'}\right),
\end{equation}

\noindent with $S'=10$\,mJy (fixed at a well-measured part of the flux
distribution), $N'=(490\pm104)$\,deg$^{-2}$ and $\alpha=3.0\pm0.7$.

At flux densities above 20\,mJy, the number counts from this survey are
complemented by the equivalent measurements from {\it Herschel} surveys, which
survey wider areas at 500$\mu$m to shallower depths, and so find the rarer,
bright sources (nearby galaxies, extremely luminous distant sources and
gravitationally lensed galaxies) that are not present in our map (Clements et
al.\ 2010; Negrello et al.\ 2010). We focus on two {\it Herschel} surveys;
HerMES, which has obtained confusion limited maps reaching a detection limit
of $S_{500}\approx20$\,mJy (Oliver et al.\ 2012) and the {\it Herschel}-ATLAS
survey, which has mapped several hundreds of square degrees at a shallower
depth (Eales et al.\ 2010). As Fig.\ 4 shows, our 450$\mu$m counts are in
excellent agreement at $\approx$20\,mJy where the {\it Herschel} and SCUBA--2
CLS survey flux distributions meet. Below {\it Herschel}'s confusion limit the
500$\mu$m galaxy number counts have been inferred statistically, by both
stacking (B\'ethermin et al.\ 2012b) and pixel fluctuation analyses (Glenn et
al.\ 2010), again indicating consistency with the directly measured 450$\mu$m
number counts in approximately the same flux regime.

Recently, Chen et al.\ (2012) presented SCUBA--2 450$\mu$m observations of
SMGs in the field of the lensing cluster A\,370. The benefit of observing a
lensing cluster is -- provided a lens model is known -- the ability to probe
further down the luminosity function than would otherwise be possible for the
same flux limit, with faint background sources boosted by the cluster
potential. We compare the `delensed' counts of 450$\mu$m emitters derived from
12 galaxies in the field of A\,370 in Fig.\,4, indicating broad agreement with
our blank-field counts within the errors in the same flux range. After
delensing, Chen et al.\ (2012) are able to probe slightly fainter than our
catalogue, and the number counts at the 4.5\,mJy level are also consistent
with an extrapolation of our best fit Schechter function to the same limit.

\begin{figure}
\centerline{\includegraphics[width=0.38\textwidth,angle=-90]{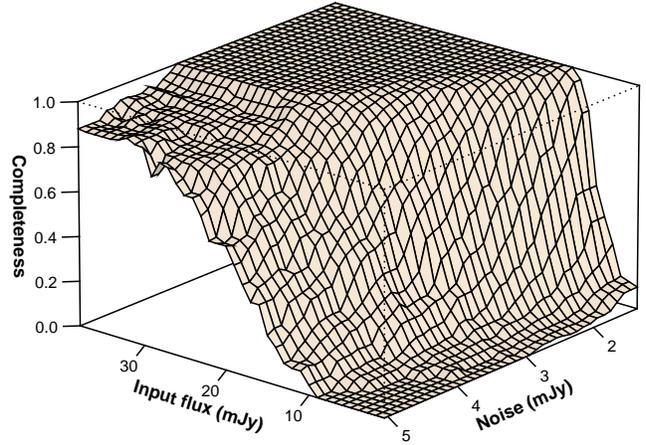}}
\caption{Completeness of the 450$\mu$m catalogue as a function of local noise
and input flux based on input-and-recovery simulations using jackknife
realisations of the map noise. Modelling the completeness as a 2-dimensional
function is required due to the radially varying sensitivity in the map
(Fig.\,\ 1). The same simulations allow us to estimate the difference between
true (input) flux and recovered (i.e.\ observed) flux densities across the
same parameter space, and we use this information to correct the number counts
accordingly. } \end{figure}

\begin{figure*}
\centerline{\includegraphics[height=\textwidth,angle=-90]{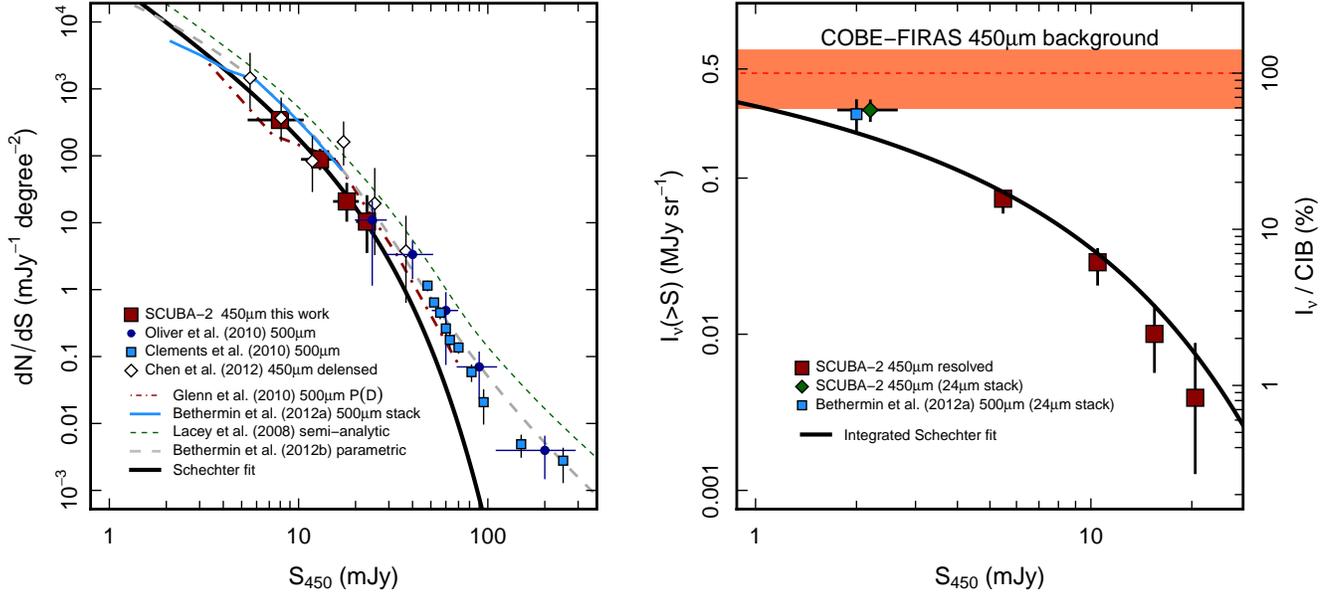}}
\caption{(left) Differential number counts of galaxies detected at 450$\mu$m
(error bars are derived from Poisson statistics). We compare the 450$\mu$m
counts to those measured recently by Chen et al. (2012) and by {\it Herschel}
at 500$\mu$m (B\'ethermin et al.\ 2012a; Glenn et al.\ 2010; Oliver et al.\
2010; Clements et al.\ 2010), and to the predictions of a numerical model of
galaxy formation (Baugh et al.\ 2005; Lacey et al.\ 2008) and a parametric
model of the evolution of the infrared luminosity density of the Universe
(B\'ethermin et al.\ 2012b). We fit the measured 450$\mu$m counts with a
Schechter function; the reason it fails to reproduce the shape of the
500$\mu$m number counts (and those predicted by the models) at bright fluxes
is because in this flux regime the counts have significant contributions from
(a) bright local star-forming galaxies and (b) distant galaxies boosted to
high observed flux by gravitational lensing; the SCUBA--2 map is too small to
adequately sample these populations. Note -- no 450$\mu$m/500$\mu$m colour
correction has been made to the 500$\mu$m data. (right) Integrated surface
brightness of 450$\mu$m emitters relative to the CIB measured by {\it
COBE}--FIRAS at 450$\mu$m (Fixsen et al.\ 1998). The directly measured number
counts are well-fitted by a Schechter function, the extrapolation of which
agrees with the CIB derived from a stack of 24$\mu$m-emitting galaxies not
individually detected in the SCUBA--2 map. Thus, we directly resolve
16$\pm$7\% of the CIB measured by {\it COBE}--FIRAS into galaxies, with an
additional $\approx$40\% contributed by 24$\mu$m-emitting galaxies not
formally detected at 450$\mu$m. We project that 100\% of the CIB at 450$\mu$m
is recovered at a flux density of $S_{450}>0.4$\,mJy.} \end{figure*}

\subsection{Resolving the 450${\bf \mu}$m background light}

What fraction of the CIB at 450$\mu$m have we resolved into galaxies? The
integrated flux density of point sources detected at 450$\mu$m (corrected for
completeness) is $I_\nu(450\mu{\rm m}) = (7.4\pm 0.7)\times
10^{-2}$\,MJy\,sr$^{-1}$. The absolute intensity of the CIB at 450$\mu$m
measured by {\it COBE}--FIRAS is $I_\nu(450\mu{\rm m}) =
0.47\pm0.19$\,MJy\,sr$^{-1}$, thus we have directly resolved $16\pm7$\% of the
CIB at 450$\mu$m (the uncertainty is dominated by the {\it COBE}--FIRAS
measurement; Fixsen et al.\ 1998). For comparison, the deepest {\it Herschel}
surveys have directly resolved 5--6\% of the CIB at 500$\mu$m (Oliver et al.\
2010; B\'ethermin et al.\ 2012a). We show the integrated brightness of the
450$\mu$m emitters, relative to the absolute intensity of the CIB in Fig.\,\
4.

To measure the contribution to the CIB at 450$\mu$m by galaxies not formally
detected in the SCUBA--2 map, but which are known to be  infrared--bright
galaxies, we stack the map at the position of 1600 galaxies selected from a
catalogue generated from the {\it Spitzer}-COSMOS MIPS 24$\mu$m image of the
same region (Sanders et al.\ 2007). First, we remove point sources from the
450$\mu$m map, using a point spread function (PSF) constructed by averaging
the two dimensional profiles of sources detected at $>$7$\sigma$. This PSF was
then normalised to the flux of each individual source in our catalogue, and
subtracted from the map. This yields a residual map where the only flux (in
addition to that of noise) is contributed by sources not in our catalogue. The
450$\mu$m is then stacked at the position of the 24$\mu$m sources, averaging
the flux with a weight equivalent to the inverse of the variance of the map at
each position.

The average 450$\mu$m flux density of 24$\mu$m sources with mean 24$\mu$m flux
$\langle S_{24}\rangle=0.19$\,mJy is $\langle S_{450}\rangle =
2.2\pm0.4$\,mJy. The resulting contribution to the 450$\mu$m background is
$0.20\pm0.04$\,MJy\,sr$^{-1}$ (the uncertainty is $\sigma/\sqrt N$, with
$\sigma$ the standard deviation in the stack and $N$ the sample size). A
simple simulation was performed to test whether the stacking methodology
described above produces unbiased estimates of the submillimetre flux. The
residual flux map was inverted (by multiplying by $-1$) and simulated sources
were inserted using the derived PSF as a model, with the input fluxes of the
fake sources set to $S_{450}=10S_{24}$ up to a maximum of $S_{450}=5$\,mJy.
The positions were set to the 24$\mu$m catalogue positions, rotated 90 degrees
about the map centre, thus preserving clustering information. The stacking
procedure was then repeated as for the real catalogue. The mean input flux was
$S_{450}=1.8$\,mJy per source, and the recovered stacked flux was
$S_{450}=1.0\pm0.5$\,mJy. The recovered flux is slightly low compared to the
input flux at the 1.5$\sigma$ level, however this does not affect our
conclusions, given the uncertainties in the 450$\mu$m flux calibration and the
absolute measured value of the CIB at 450$\mu$m.

Excluding those detected as bright point sources, the 24$\mu$m-selected
galaxies contribute $(2.0\pm0.4)\times10^{-1}$\,MJy\,sr$^{-1}$, or $42\pm19$\%
of the CIB at 450$\mu$m. Therefore, in addition to the directly detected
sources, in total we can account for $58\pm20$\% of the CIB at 450$\mu$m using
the SCUBA--2 map. Note that our stacked value is in good agreement with the
background derived from a stack of 24$\mu$m-emitters in {\it Herschel}
500$\mu$m images (B\'ethermin et al.\ 2012b), and is also consistent with the
intensity expected from an extrapolation of a Schechter function fit to the
directly measured number counts (Fig.\,4).


\begin{table}

 \caption{Number counts of 450$\mu$m-selected galaxies. $N$ indicates the raw number of galaxies in each bin ($\delta
S_{450}=5$\,mJy), and the completeness ($\mathcal{C}$) and de-boosting
($\mathcal{B}$) corrections represent the mean corrections for galaxies in
each bin (note that each galaxy is de-boosted individually, with the
correction increasing for lower flux densities). Uncertainties in the
differential counts are the 1$\sigma$ confidence range assuming Poisson
statistics (Gehrels\ 1986).}

\begin{center}
	\begin{tabular}{cccccc}
		\hline
		$S_{450}$ & $N$ & $dN/dS$ & $N(>$$S')$$^{\rm a}$ & $\langle\mathcal{C}\rangle$$^{\rm b}$ & $\langle\mathcal{B}\rangle$$^{\rm c}$ \cr
		(mJy) & & (mJy$^{-1}$\,deg$^{-2}$) & (deg$^{-2}$) &  & \cr 
		\hline
		$8.0$ & $41$ & $343.0^{+62.6}_{-53.3}$ & $2313.4^{+339.7}_{-297.7}$ & $1.6$ & $1.1$ \cr
		$13.0$ & $13$ & $88.5^{+32.3}_{-24.2}$ & $598.5^{+172.4}_{-136.0}$ & $1.1$ & $1.1$ \cr
		$18.0$ & $4$ & $20.8^{+16.8}_{-9.9}$ & $155.9^{+94.7}_{-61.7}$ & $1.0$ & $1.0$ \cr
		$23.0$ & $2$ & $10.4^{+14.2}_{-6.7}$ & $52.0^{+71.0}_{-33.5}$ & $1.0$ & $1.0$ \cr

		\hline
		\multicolumn{6}{l}{$^{\rm a}$$S'$\,corresponds to the lower edge of the bin, i.e.\ $(S_{450}-2.5)$\,mJy }\cr
		\multicolumn{6}{l}{$^{\rm b}$\,average completeness correction applied}\cr
		\multicolumn{6}{l}{$^{\rm c}$\,average flux de-boosting correction applied}\cr
	\end{tabular}
	
\end{center}
\end{table}

\section{Discussion}

We compare our results to the phenomenological model of B\'ethermin et al.\
(2012b), who use a `backwards evolution' parameterisation of the the infrared
luminosity density (as traced by dusty star-forming galaxies; see also Lagache
et al.\ 2004). The B\'ethermin et al. (2012b) model assumes that the star
formation modes of galaxies can be either described as `main sequence' (i.e.\
SFR scales with stellar mass) or `starburst', with spectral energy
distributions defined by the latest stellar synthesis template libraries. The
evolution of the luminosity functions of these two populations integrated over
cosmic history provides good fits to the observed number counts of galaxies at
24, 70, 100, 160, 250, 350, 500, 850, 1100$\mu$m and 1.4\,GHz (as well as
integrated observables such as the evolution of the volume averaged star
formation rate and cosmic infrared background). Here we confirm that the
number counts of 450$\mu$m emitters predicted by the model is also in good
agreement with the measured 450$\mu$m number counts in the flux range probed
by our SCUBA--2 survey.

We also compare the measured counts to the {\sc galform} semi-analytic model
of galaxy formation (Cole et al.\ 2000; Baugh\ 2006; Lacey et al.\ 2008;
Almeida et al.\ 2011). This prescription predicts the formation and evolution
of galaxies within the $\Lambda$\,CDM model of structure formation (Springel
et al.\ 2005), and includes the key physics of the galaxy formation (and
evolution) process: radiative cooling of gas within the dark matter halos,
quiescent (by which we mean non-burst driven) star formation in the resultant
discs, mergers, chemical enrichment of the stellar populations and
intergalactic medium and feedback from supernovae and active galactic nuclei.
As Fig.\ 4 shows, the numerical model slightly over-predicts the abundance of
450$\mu$m emitters in the flux range probed. Nevertheless, the reasonable
agreement between the shape of the counts predicted by {\sc galform} and the
data is encouraging for models of galaxy formation that aim to reproduce the
full range of emission processes of galaxies at long wavelengths.

The 8\,arcsec resolution of the 450$\mu$m SCUBA--2 map allows us to accurately
identify the optical/near-infrared counterparts of the SMGs, and we have
identified the most likely counterpart to the majority of 450$\mu$m sources in
our sample (I.\ G.\ Roseboom et al.\ 2012 in prep). The wealth of legacy data
available in the COSMOS field then provides the means to estimate the redshift
distribution of the population. We have used 13 bands of optical/near-infrared
photometry, including CFHT {\it ugri}, Subaru SuprimeCam {\it z'}, VISTA {\it
YJHK}, {\it HST} F125W and F160W and {\it Spitzer} IRAC [3.6] and [4.5] to
evaluate the photometric redshifts of all the identified galaxies (the typical
1$\sigma$ uncertainty based on the confidence level of the template fit is
$\delta z=0.16$). The redshift distribution is shown in Fig.\,\ 5, indicating
that the majority of our sample lie at $z<3$, with a mean redshift of $\langle
z \rangle=1.3$ (a full analysis of the source identification and redshift
distribution is to be presented in Roseboom et al.\, 2012 in prep). This is a
clear indication that the 450$\mu$m selection is probing a lower redshift
population than previous 850$\mu$m selected samples, which have typical
redshifts of $\langle z \rangle=2.2$ (e.g.\ Chapman et al.\ 2005; Wardlow et
al.\ 2010). The shape of the redshift distributions predicted both by the
phenomenological model and numerical model described above (for galaxies at
the same flux limit) are also in good agreement with the measured
distribution; both models predict little contribution from galaxies at $z>3$
(although a high redshift `tail' is present in both models).

\begin{figure}
\centerline{\includegraphics[width=0.45\textwidth,angle=-90]{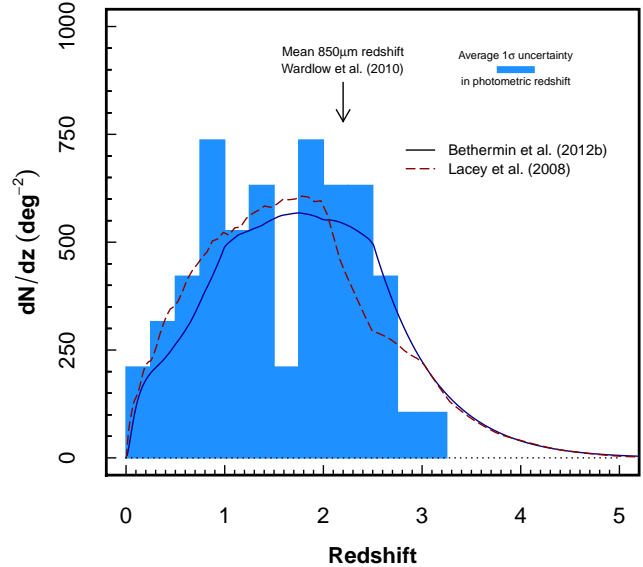}}
\caption{Redshift distribution of 450$\mu$m-selected SMGs in our sample (not
corrected for completeness), derived from 13-band photometric redshift
estimates (\S3.4). The average redshift of the sample is $\langle z
\rangle$=1.3, and the vast majority of the 450$\mu$m-selected SMGs lie at
$z<3$. For comparison, the average redshift of SMGs selected at 850$\mu$m is
$z\approx2.2$ (Wardlow et al.\ 2010), indicating the efficacy at which the
450$\mu$m selection samples a population of SMGs at lower redshift and
therefore an important complement to any census of the dusty Universe. We
compare the shape of the redshift distribution to the models of B\'ethermin et
al.\ (2012b) and Lacey et al.\ (2008) shown in Fig.\,4 for galaxies with
$S_{450}>5$\,mJy (we have area-normalised both model distributions, since the
observed redshift distribution contains no completeness correction). The
average redshift and shape of both model distributions is in good agreement
with observations, suggesting that -- at this flux limit -- there is little
contribution from galaxies at $z>3$.} \end{figure}

Assuming the directly detected sources representing 16$\pm$7\% of the CIB at
450$\mu$m are star-forming galaxies at $\langle z \rangle=1.3$, then their
total (rest-frame 8--1000$\mu$m) luminosities are in the ultraluminous class,
$L_{\rm IR}\approx 1.1\times 10^{12}{\rm L_\odot}$ (Chary \& Elbaz\ 2001). If
the galaxies contributing to the 24$\mu$m stack described in \S3.3 lie in the
same redshift range as the directly detected galaxies, then the majority
(60\%) of the CIB at 450$\mu$m is emitted by galaxies with $L_{\rm
IR}>3.6\times10^{11}{\rm L_\odot}$. This is broadly consistent with the
picture that at $z\approx1$, the star formation rate budget of the Universe is
dominated by galaxies in the LIRG class, with star formation rates of order
10\,${\rm M_\odot}$\,yr$^{-1}$ (Dole et al.\ 2006; Rodighiero et al.\ 2010;
Magnelli et al.\ 2011).

An extrapolation of the Schechter function fit to the directly measured number
counts (which agrees well with the background at $S_{450}\approx2$\,mJy,
derived from the stack of 24$\mu$m sources), implies that 100\% of the CIB at
450$\mu$m should be recovered at $0.1<S_{450}<1.4$\,mJy (the range accounting
for the 1$\sigma$ uncertainty of the absolute measured background; Fixsen et
al. 1998), close to the SCUBA--2 confusion limit. If the galaxies responsible
for this emission are at similar redshifts to the current 450$\mu$m sample
(but below the sensitivity of the map and not contributing to the 24$\mu$m
stack), then the majority of the remaining $\approx$40\% of the CIB at
450$\mu$m is likely to be emitted by galaxies with $L_{\rm
IR}<1.3\times10^{11}{\rm L_\odot}$, implying galaxies star formation rates of
a few tens of Solar masses per year. However, we cannot as yet rule out what
fraction of the remaining CIB light might be emitted by faint 450$\mu$m
emitters at higher redshifts; note that a galaxy with $S_{450}\approx 2$\,mJy
at $z>2$ has a typical luminosity of $L_{\rm IR}>5.5\times10^{11}{\rm
L_\odot}$ (Chary \& Elbaz\ 2001), again indicating the importance of
LIRG-class galaxies in the cosmic infrared budget. Characterizing the high
redshift tail of the 450$\mu$m population is an important next step.

\section{Summary}

The SCUBA--2 camera on the 15\,m JCMT repesents the state-of-the-art in
panoramic submillimetre imaging, and has recently begun scientific
observations in earnest. In this paper we have presented results from the
first deep, blank-field cosmological map at 450$\mu$m
($\sigma_{450}=1.3$\,mJy); part of the SCUBA--2 Cosmology Legacy Survey, the
largest of the seven JCMT Legacy Surveys. Using a 450\,$\mu$m map of the
well-studied extragalactic COSMOS/CANDELS field, we have

\begin{enumerate}

\item made the first unbiased, blank-field determination of the number counts
of galaxies at 450$\mu$m, at a flux density limit of $S_{450} > 5$\,mJy. This
probes below the confusion limit of {\it Herschel}, complementing the number
counts measured at fluxes above 20\,mJy over wider areas in major {\it
Herschel} submillimetre surveys;

\item measured the contribution of these galaxies to the cosmic infrared
background at 450$\mu$m: we resolve 16\% of the CIB into individual galaxies.
The ability of SCUBA--2 to `pin-point' the galaxies responsible for the
emission of the CIB is a critical step in understanding the properties of the
galaxies that are forming the majority of stars in the Universe at this epoch;

\item an additional $\approx$40\% of the CIB can be recovered in the SCUBA--2
map by stacking {\it Spitzer} MIPS-detected 24$\mu$m emitters. Using this
analysis we estimate that the majority ($\approx$60\%) of the CIB at 450$\mu$m
is emitted by galaxies with $S_{450}>2$\,mJy;

\item a preliminary analysis of the redshift distribution of the 450$\mu$m
emitters (based on high-quality photometric redshifts available for this
field) imply that the typical redshift of galaxies with $S_{450}>5$\,mJy is
$\langle z\rangle=1.3$, with the majority lying at $z<3$. The typical
luminosity of galaxies in our sample are estimated to be in the ultraluminous
class, with $L_{\rm IR}>10^{12}{\rm L_\odot}$. If the galaxies contributing to
the statistical stack of 24$\mu$m emitters described above are at a similar
redshift, then we project that the majority of the CIB at 450$\mu$m is emitted
by `LIRG' class galaxies with $L_{\rm IR}>1.3\times 10^{11}{\rm L_\odot}$.

\end{enumerate}

\noindent These are the first results of the S2CLS. The final goal of the
survey will be to map a quarter of a square degree to $\sigma_{450}=1.2$\,mJy,
and a wider, ten square degree area to $\sigma_{850}=1.5$\,mJy, yielding
$>$10$^4$ SMGs with which to (a) determine the submillimetre luminosity
function and its evolution over cosmic time; (b) search for the rarest, most
luminous SMGs at high-{\it z}; (c) resolve the 450$\mu$m background; (d)
accurately measure the clustering properties of SMGs and determine their
typical environments (including rare protoclusters) and (e) build-up the large
samples required to properly relate SMGs to other star-forming
(ultraviolet/near-infrared selected) populations at $z\sim2$ and thus gain
further insight into SMGs' role in the overall history of galaxy evolution.

\section*{Acknowledgements}

J.E.G. is supported by a Banting Fellowship, administered by NSERC. J.S.D.
acknowledges the support of the European Research Council via the award of an
Advanced Grant, and the support of the Royal Society via a Wolfson Research
Merit award. M.M. and I.G.R. acknowledge the support of STFC. The authors
thank M. B\'ethermin for providing data on the 500$\mu$m stacking and
parametric model, and J. Dempsey for advice on the SCUBA--2 calibration. It is
also a pleasure to thank the JCMT telescope operators J. Hoge, J. Wouterloot
and W. Montgomerie, without whom these observations would not be possible. The
James Clerk Maxwell Telescope is operated by the Joint Astronomy Centre on
behalf of the Science and Technology Facilities Council of the United Kingdom,
the Netherlands Organisation for Scientific Research, and the National
Research Council of Canada. Additional funds for the contruction of SCUBA-2
were provided by the Canada Foundation for Innovation. {\it Herschel} is an
ESA space observatory with science instruments provided by European-led
Principal Investigator consortia and with important participation from NASA.
This work is based in part on observations made with the {\it Spitzer Space
Telescope}, which is operated by the Jet Propulsion Laboratory, California
Institute of Technology under a contract with NASA.

\label{lastpage}

\end{document}